\documentclass[leqno]{statsoc}

\usepackage[latin1]{inputenc}
\usepackage{amsmath,amssymb,amsfonts,latexsym,mathrsfs}
\usepackage{epsfig,graphicx,rotating}
\usepackage{fullpage,stmaryrd}
\usepackage{fancyhdr,pstricks}
\usepackage{natbib}
\bibpunct{(}{)}{,}{a}{,}{,}
\newcommand\btheta{\mathbf{\theta}}
\newcommand\bp{\mathbf{p}}

\title[Discussions on ``Riemann manifold Langevin and Hamiltonian Monte Carlo methods"]{Discussions on
the Read Paper by Girolami and Calderhead ``Riemann manifold Langevin and Hamiltonian Monte Carlo
methods" read to the Society on October 13th, 2010} \author{Simon Barthelm\'e${}^1$, 
Magali Beffy${}^{2,3}$, 
Nicolas Chopin${}^2$, Arnaud Doucet${}^4$, Pierre Jacob${}^{2,3}$, Adam M.~Johansen${}^5$, 
Jean-Michel Marin${}^6$, and Christian P. Robert${}^{2,4}$}
\address{${}^1$BCCN (TU Berlin),~${}^2$CREST, Paris,
~${}^3$Universit\'e Paris--Dauphine, CEREMADE,
~${}^4$Dept. Statistics and Computer Science, UBC, Vancouver,
~${}^5$Department of Statistics, University of Warwick,
and
~${}^6$Universit\'e de Montpellier 2, I3M}
\begin{document} 

\maketitle

\begin{abstract}
This is a collection of discussions of `Riemann manifold Langevin and Hamiltonian Monte Carlo
methods" by Girolami and Calderhead, to appear in the {\em Journal of the Royal Statistical 
Society}, Series B.
\end{abstract}

\makeatother

\section{Connections with optimisation {\em (S.~Barthelm\'e and N.~Chopin)}}

One of the many things we like about this paper is that it forces
us to change our perspective on Metropolis-Hastings. We may not the
only ones with the toy example of a bivariate, strongly correlated,
Gaussian distribution imprinted in our brain. This example explains
well why taking correlations into account is important. However, one
often forgets that, contrary to the Gaussian example, the curvature
of the log target density may be far from constant, which justifies
a \emph{local} calibration of HM strategies. The authors give compelling
evidence that local calibration may lead to strong improvements in
large-dimensional problems.

There are two ways to understand these results. One of them, put forward
in this paper, stems from the information geometry perspective: the
parameter space is endowed with a metric defined by $G(\theta)$,
which turns the posterior distribution into a density over a manifold.
The general MMALA algorithm based on a diffusion over that manifold
is a beautiful mathematical device, but it is not immediately apparent
how this leads to improved (relative) MCMC performance. A different
viewpoint proceeds from optimisation: MMALA performs better because
it uses a better local model of the posterior density. 

As often pointed out, the Langevin proposal is a noisy version of
a gradient ascent step. Similarly, the simplified MMALA step is a
noisy version of a (quasi-)Newton step, in which the Hessian is replaced
with the Fisher information matrix, an idea known as Iteratively Reweighted
Least-Squares in the literature on Generalised Linear Models. It is
worth emphasising the fact that the simplified versions, which just
relies on these local curvature ideas, but do not require third derivatives,
do better in terms of relative efficiency (not to mention in terms
of human computation time!). 

This suggests two avenues for further research. First, many optimisation
methods have been developed that only require evaluating the gradient.
This may be more convenient from the practitioner's point of view,
and it also proves more effective whenever computing Hessian matrices
is expensive. Methods, such as the BFGS or Barzilai-Borwein, approximate
the Hessian locally from the previous $k$ iterations. Our preliminary
experiments indicate that these methods may reduce the correlation
in MCMC chains.

The second point is that the auxiliary Gaussian distribution is merely
a choice imposed by the physical interpretation of the Hamiltonian.
Do the authors have any intuition on what would be the optimal auxiliary
distribution?

\section{Multiple potentials {\em (M.~Beffy and C.P.~Robert)}}
The paper gives a very clear geometric motivation for the use of an Hamiltonian representation.
As such, it suggests for an immediate generalisation by extending the Hamiltonian dynamic to a more general dynamic on 
the level sets of 
\begin{align*}
\mathscr{H}(\btheta,&\mathbf{p}_1,\mathbf{p}_2,\ldots,\mathbf{p}_k)=
-\mathcal{L}(\btheta)+\frac{1}{2}\log\{(2\pi)^D|\mathbf{G}_1(\btheta)|\}
+\frac{1}{2}\mathbf{p_1}^\text{T}\mathbf{G}_1(\btheta)^{-1}\mathbf{p}_1\\
&+\frac{1}{2}\log\{(2\pi)^D|\mathbf{G}_2(\btheta)|\}+\frac{1}{2}\mathbf{p_2}^\text{T}\mathbf{G}_2(\btheta)^{-1}\mathbf{p}_2
+\ldots\\
&+\frac{1}{2}\log\{(2\pi)^D|\mathbf{G}_k(\btheta)|\}+\frac{1}{2}\mathbf{p_k}^\text{T}\mathbf{G}_k(\btheta)^{-1}\mathbf{p}_k\,,
\end{align*}
where the $\mathbf{p}_j$s are auxiliary vectors of the same dimension $D$ as $\btheta$ and the $\mathbf{G}_j(\btheta)$s are symmetric
matrices. This function is then associated with the pde's 
\begin{align*}
\dfrac{\text{d}\theta_i}{\text{dt}} &= \dfrac{\partial}{\partial p_{ij}} \mathscr{H}(\btheta,
\mathbf{p}_1,\ldots,\mathbf{p}_k) = \left\{ G_j(\btheta)^{-1} \bp_j \right\}_i\\
\dfrac{\text{d}p_{ij}}{\text{dt}} &= - \dfrac{\partial}{\partial \theta_{i}} \mathscr{H}_j(\btheta,
\mathbf{p}_1,\ldots,\mathbf{p}_k)
\end{align*}
in that those moves preserve the potential $\mathscr{H}(\btheta,\mathbf{p}_1,\ldots,\mathbf{p}_k)$ 
and hence the target distribution at all times $t$. This generalisation would allow for using a range of information matrices
$\mathbf{G}_j(\btheta)$s in parallel. The corresponding RHMC implementation is to pick one of the indices $j$ at random and to
follow the same moves as in the paper, given the separation between the different energies.

\section{Information approximations {\em (A. Doucet, P. Jacob and A.M. Johansen)}}
Congratulations to the authors for their elegant contribution.

Consider those situations in which one does not have direct access to an
appropriate metric but can obtain pointwise, simulation-based estimates of its values. For example, we might be interested in performing Bayesian
inference in general state-space Hidden Markov Models (HMM) using particle
MCMC methods \citep{andrieu2010}. In this context, we integrate out
numerically the latent variables of the model using a Sequential Monte Carlo
(SMC) scheme. A sensible metric to use is the observed information matrix
which can also be estimated using SMC\ \citep{george2010}. We discuss here
the use of such estimates in an MCMC\ context.

Assume we want to sample from a target $\pi (x)$ on $\mathcal{X}$ using the
Metropolis-Hastings (M-H) algorithm. Denote the proposal's parameters (e.g.
scale) $r\in \mathcal{R}$. Defining an extended target over $\mathcal{X}%
\times \mathcal{R}$, as $\overline{\pi }(x,r)=\pi (x)q(r|x)$ an algorithm
may be defined on $\mathcal{X}\times \mathcal{R}$ in which both $R$ and $X$
are sampled.

At iteration $n+1$ draw $X^{\star }\sim s(\cdot |x_{n},r_{n})$ and $R^{\star
}\sim q(\cdot |x^{\star })$. Accept this proposal with the standard MH
acceptance probability on the extended space 
\begin{align*}
\alpha (x_{n},r_{n};x^{\star },r^{\star })=& 1\wedge \frac{\overline{\pi }%
(x^{\star },r^{\star })}{\overline{\pi }(x_{n},r_{n})}\cdot \frac{%
s(x_{n}|x^{\star },r^{\star })q(r_{n}|x_{n})}{s(x^{\star
}|x_{n},r_{n})q(r^{\star }|x^{\star })} \\
=& 1\wedge \frac{\pi (x^{\star })}{\pi (x_{n})}\cdot \frac{s(x_{n}|x^{\star
},r^{\star })}{s(x^{\star }|x_{n},r_{n})}.
\end{align*}%
Hence it is not necessary to be able to evaluate $q$, even pointwise,
provided that it can be sampled from. The resulting transition is reversible
on the extended space and admits $\pi $ as a marginal of its invariant
distribution. This simple result is well-known: see Besag (\citeyear[Appendix 1]{besag95}).

The MMALA, with metric tensor obtained by sampling, may be justified using
precisely the same argument: A proposal of the form of (10), may be
implemented with a sampled estimate of the metric tensor and such gradients
as are required (objects which can be obtained readily in settings of
interest, such as HMMs); the extended space construction above holds with $x
= \theta; r=(G,\nabla G)$ and the acceptance probability remains of the same
form; the constant curvature proposal may be implemented without the need
for estimates of $\nabla G$ with $x=\theta$ and $r=G$.

The HMC variant of the same is not trivial. As each step of the implicit
integrator requires access to the value of the metric at several
(implicitly-defined) points, direct application of the above principles does
not appear possible. However, more subtle approaches can be employed. In
particular one could consider trying to approximate the metric using the
expectation of a function with respect to a probability measure independent
of $x$ and using common random variates from this measure during an HMC
update.

\section{On some examples {\em (J.-M.~Marin and C.P.~Robert)}}
This paper is a welcome addition to the recent MCMC literature and the authors are
to be congratulated for linking together the two threads that are the Langevin modification of the
random walk Metropolis--Hastings algorithm and the Hamiltonian acceleration.  Overall, trying to
take advantage of second order properties of the target density $\pi(\theta)$, just like the Langevin improvement takes
advantage of the first order \citep{roberts:tweedie:1995,stramer:tweedie:1999a,stramer:tweedie:1999b}
is a natural idea which, when implementable, can obviously speed up
convergence. This is the Langevin part, which may use a fixed metric $\mathbf{M}$ or a local metric defining a
Riemann manifold, $\mathbf{G}(\theta)$. Obviously, this requires that the derivation of an appropriate (observed or expected)
information matrix $\mathbf{G}(\theta)$ is feasible up to some approximation level. Or else that authoritative enough
directions are given about the choice of an alternative $\mathbf{G}(\theta)$.

While the logistic example used in the paper mostly is a toy problem (where importance sampling works extremely well, as shown in 
\citealp{marin:robert:2010}), the stochastic volatility model is more challenging and the fact that the Hamiltonian scheme
applies to the missing data (volatility) as well as to the three parameters of the model is quite interesting. We would thus
welcome more details on the implementation of the algorithm in such a large dimension space. We however
wonder at the appeal of this involved scheme when considering that the full conditional distribution of the volatility can be
simulated exactly.

\section{Moving away from continuous time {\em (C.P. Robert)}}
This paper is an interesting addition to recent MCMC literature and I am eager to see how the 
community is going to react to this potential addition to the MCMC toolbox. I am however wondering about the
impact of the paper on MCMC practice. Indeed, while the dynamic on the level sets of 
$$ 
\mathscr{H}(\theta,\mathbf{p})=
-\mathcal{L}(\theta)+\frac{1}{2}\log\{(2\pi)^D|\mathbf{G}(\theta)|\}+\frac{1}{2}\mathbf{p}^\text{T}\mathbf{G}(\theta)^{-1}\mathbf{p}\,,
$$ 
where $\mathbf{p}$ is an auxiliary vector of dimension $D$,
is associated with Hamilton's equations, in that those moves preserve the potential
$\mathscr{H}(\theta,\mathbf{p})$ and hence the target distribution at all times $t$, I argue that
the transfer to the simulation side, i.e.~the discretisation part, is not necessarily useful, or at least that it
does not need to be so painstakingly reproducing the continuous phenomenon. 

\begin{figure}[hb]
\includegraphics[width=.95\textwidth]{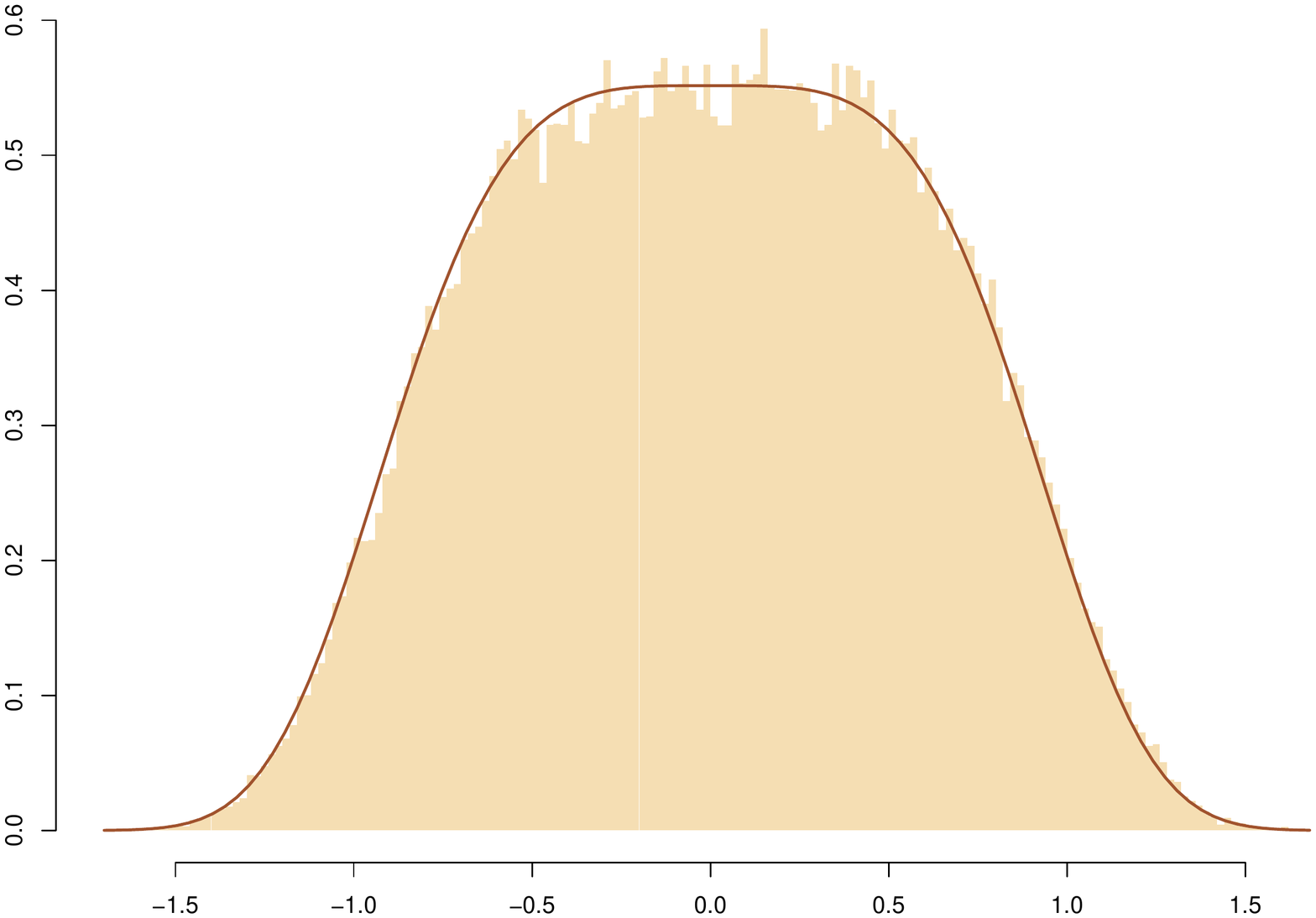}
\includegraphics[width=.95\textwidth]{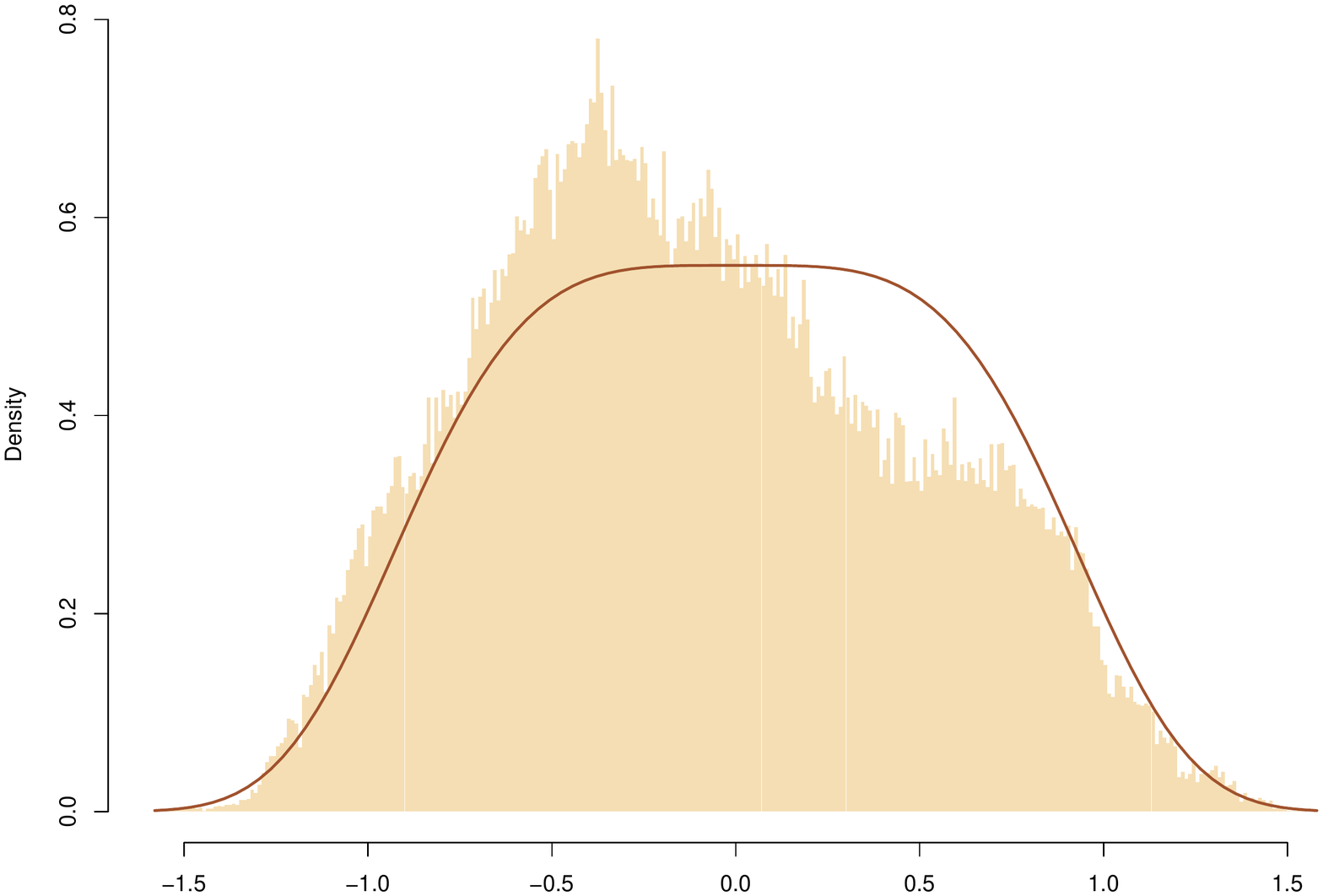}
\caption{\label{fig:t4T}Comparison of the fits of discretised Langevin diffusions to the target $f(x)\propto\exp(-x^4)$ when using
a discretisation step $\sigma^2=.01$ {\em (left)} and $\sigma^2=.0001$ {\em (right)}, after $T=10^7$ steps. This comparison illustrates
the need for more time steps when using a smaller discretisation step.}
\end{figure}

In a continuous time-frame, the purpose of the auxiliary vector $\mathbf{p}$
is clearly to speed up the exploration of the posterior surface by taking advantage of the additional energy it
provides. In the discrete-time universe of simulation, on the one hand, the
fact that the discretised (Euler) approximation to Hamilton's equations are 
not exact nor available in closed form does not present such a challenge in that approximations
can be corrected by a Metropolis-Hastings step, provided of course all terms in the Metropolis-Hastings ratio are available. On
the other hand, the continuous-time (physical or geometric) analogy at the core of the Hamiltonian may be unnecessary
costly when trying to carry a physical pattern in a discrete (algorithmic) time. MCMC algorithms are not set to work in
continuous time and therefore the invariance and stability properties of the continuous-time process that motivates the
method do not carry to the discretised version of the process. For one thing, the (continuous) time unit has no
equivalent in discrete time. Therefore, the dynamics of the Hamiltonian do not tell us how long the discretised version
should run, as illustrated on Figure \ref{fig:t4T}.  As a result, convergence issues (of the MCMC algorithm) should not
be impacted by inexact renderings of the continuous-time process in discrete time. For instance, when considering the
Langevin diffusion, the corresponding Langevin algorithm could as well use another scale $\eta$ for the gradient than
the one $\tau$ used for the noise, i.e. 
$$
y=x^t+\eta\nabla\pi(x)+\tau\epsilon_t
$$
rather than a strict Euler discretisation where $\eta=\tau^2/2$.
A few experiments run in \citeauthor{robert:casella:1999} (1999, Chapter 6, Section 6.5) showed that using a different scale 
$\eta$ could actually lead to improvements, even though we never pursued the matter any further.

\end{document}